# Directionally asymmetric self-assembly of cadmium sulfide nanotubes using porous alumina nanoreactor – need for chemo-hydrodynamic instability at the nanoscale

Arthur Varghese* and Shouvik Datta**

Department of Physics, Indian Institute of Science Education and Research,

Pune – 411021, Maharashtra, India.

## ABSTRACT

We explore nanoscale hydrodynamical effects on synthesis and self-assembly of cadmium sulfide nanotubes oriented along one direction. These nanotubes are synthesized by horizontal capillary flow of two different chemical reagents from opposite directions through nanochannels of porous anodic alumina which are used primarily as nanoreactors. We show that uneven flow of different chemical precursors is responsible for directionally asymmetric growth of these nanotubes. On the basis of structural observations using scanning electron microscope, we argue that chemo-hydrodynamic convective interfacial instability of multi-component liquid-liquid reactive interface is necessary for sustained nucleation of these CdS nanotubes at the edges of these porous nanochannels over several hours. However, our estimates clearly suggest that classical hydrodynamics cannot account for the occurrence of such instabilities at these small length scales. Therefore, we present a case which necessitates further investigation and understanding of chemo-hydrodynamic fluid flow through nanoconfined channels in order to explain the occurrence of such interfacial instabilities at nanometer length scales.

# I. INTRODUCTION

Ordered nanostructures are expected to have interesting physical properties and can be used in a wide range of potential applications in nanoelectronic devices. Usually, solution based wet chemical techniques are used to grow many such ordered nanostructures. However, physical understanding of nanoscale hydrodynamics in the nanofabrication process is generally overlooked. One simple way to fabricate an array of such ordered nanowires is to use nanoporous alumina as a nanotemplate. Self assembled nanoporous alumina [1,2] are usually prepared by two step anodization of aluminum metal. Such nanotemplates of anodized aluminum oxide (AAO) have been widely used to synthesize nanowires of a large variety of materials inside its columnar nanochannels by chemical deposition [3], electrochemical deposition [4] and physical deposition [5] techniques etc. Here we report a particular nanofabrication route by employing free standing nanoporous alumina membrane as a collection of '*nano chemical reactor*' to synthesize and self assemble ordered arrays of cadmium sulfide (CdS) nanowires and then finally discuss the need for chemo-hydrodynamic instabilities in this nanofabrication process.

CdS is a group II-VI compound semiconductor with a bulk band gap of ~2.4 eV. As such, study of nanofluidic flow through such confined channels is widely reported in the literature but hydrodynamics of two different fluids (chemical reagents) mixing and reacting in a nanoconfined space to produce nanostructures are not well studied. In our work, chemical reaction and subsequent nucleation of CdS nanowires actually takes place within a nanoconfined space mainly determined by the inner diameter of these AAO nanochannels. Directionally asymmetric growth of CdS nanowires oriented along one direction is observed during such nanoreactor based chemical synthesis. These nanowires are imaged by scanning electron microscopy (SEM) and characterized by localized energy dispersive X-Ray spectroscopy (EDS) studies. We argue that formation of electrical double layer (EDL) is not sufficient enough to explain any ion-selective conventional nanofluidic flow for AAO pore diameters more than 10nm at the current level of concentration of chemical reagents used in this work. In the next few paragraphs, we also talk about important structural observations from SEM imaging and discuss different hydrodynamic mechanisms behind the formation of such CdS nanotube outgrowths. We present these structural evidences to suggest that it requires chemically triggered hydrodynamic instabilities as the main triggering mechanism for prolonged nucleation of such nanowires. However, we also illustrate that standard estimate of dimensionless numbers used to quantify the onset of such instabilities is not significant enough to explain such nucleation process. This certainly shows that fresh understanding of the origin of such chemo-hydrodynamic instabilities at the nanoscale is essential.

## II. NANOREACTOR BASED SYNTHESIS AND CHARACTERIZATIONS of CdS NANOTUBES

Wet chemical syntheses were carried out using a two chamber cell as shown in Fig. 1(a). Chambers are separated by few microns thick nanoporous alumina membrane. Further details of synthesis procedures can be found in the supplemental material [6]. Dilute solutions of different chemical reagents (e.g 0.1 M of $Na_2S$ & $CdCl_2$) were kept in respective chambers on opposite sides of AAO nanochannels. These two chemical reagents could only mix and react by flowing horizontally through these nanochannels of AAO membrane [7-10]. We carefully control the heights of these two liquid reagents to prevent any significant pressure difference on both side of AAO membrane. In contrast to earlier reports, here we observe that after few hours of chemical reaction in this AAO nanoreactor, CdS nanowires [Fig. 1(b)] grow horizontally and extend by nearly a micron length *only on one side* [Fig. 2] of the porous template facing the 0.1M $Na_2S$ solution. Surprisingly, no such nanowire outgrowth is observed [Fig. 2(b)] on the other side of the porous AAO template facing the 0.1 M $CdCl_2$ solution. The inset in Fig. 2(a) shows the magnified image of these CdS nanowires grown on the $Na_2S$ side of AAO. We have also studied elemental composition of as grown CdS nanowires using EDS [insets of Fig. 2(b)]. There we hardly see any presence of elemental cadmium or sulfur in the EDS spectra measured on the $CdCl_2$ side of AAO. However, we see clear indications of the presence of both of these elements on the $Na_2S$ side of AAO. It is also obvious that elemental sulfur is present in smaller stoichiometric fraction on the surface of these 100nm wide CdS nanowires as compared to elemental cadmium. We have also characterized the crystalline nature of these nanowires and have found [6] the usual strong X-ray diffraction (XRD) peaks of 002 plane of wurtzite structure of CdS [11]. XRD peaks are broadened by poly-crystalline nature of CdS nanotubes and mean crystallite sizes of ~ 2.6 nm are estimated by the Debye-Scherrer method for all nanowires having different outside diameters ranging from 20nm to 100nm.

## III. STRUCTURAL OBSERVATIONS FROM SEM IMAGING

Figs. 3(a), 3(b) & 3(c) are showing the SEM topography of CdS nanowires grown (using 0.1M of $CdCl_2$ and 0.1M $Na_2S$ for 3 hours) on AAO nanotemplate prepared with 0.3M Oxalic acid under different anodization voltages. As expected, the average diameter of these nanowires

gradually increases with the increasing width of AAO pores fabricated under higher anodization voltages. Similar directionally asymmetric growth is also observed [Fig. (3d)] by reducing the concentration of both chemical precursors to a level of 0.005M. Increased dilution only increases time duration (~24 hours) required to reach one particular length of the nanowire. We notice - (a) the tip or the top-end of each nanowire is closed at any growth duration [Fig. 4] irrespective of AAO pore diameter and (b) tube like circular wall formations inside the porous channels on the surface of AAO template are also clearly visible in SEM images shown in Fig. 1b and also in Fig. 3b. Therefore, internally these CdS nanowires grow as hollow nanotubes and not as solid nanorods. From now on, we will refer to these nanowires as nanotubes in the remaining part of this report. Moreover, we notice that – (c) the thickness (~20nm) of these nanotube walls [Fig. 3] are not changing significantly either with increasing diameter of the AAO nanopores or with changing concentration of the chemical precursors. Later, we will argue why this wall thickness may depend on some characteristic length scale over which two chemical reagents mix during growth nucleation. It is also clear from SEM images that the nucleation and subsequent growth processes of these CdS nanotubes happen concurrently as evident from the near identical length of these nanotubes for each individual sample.

# IV. DISCUSSION ON HYRODYNAMICAL EFFECTS IN THE NANOSCALE GROWTH PROCESS.

## A. Is it a typical nanofluidic flow governed by Debye length and electric double layer?

Based on these SEM images, it seems most likely that the directionally asymmetric formation of CdS nanotubes progress as the cadmium containing ions (like $Cd^{++}$ or $[CdCl_4]^{-2}$ complex) flow through these nanoporous channels, reach the other end of AAO and then react with sulfur containing ions (like $S^{-2}$ or $HS^-$) only near the exit ends of these nanochannels on $Na_2S$ side of the AAO template. Moreover, any pore filling columnar growth of CdS inside the AAO nanochannels could have clogged the outgrowth of these nanowires by preventing the necessary mixing of the chemical precursors. Flow of these sulfur containing anions through these nanochannels are possibly hindered as compared to that of cadmium containing ions due to the difference in pH of both precursor solutions. This will be elaborated in the next Sec. IV B. Otherwise, similar outgrowths of CdS nanowires should also be present on the side of AAO facing the $CdCl_2$ precursor. At this stage, we wish to estimate the usual Debye length [12] of these electrolytic solutions which is

$$\lambda_{Debye} = \sqrt{\frac{\varepsilon \varepsilon_0 k_B T}{2nz^2 e^2}} \qquad (1)$$

, where $\varepsilon$ is the relative dielectric constant of a ionic fluid, $\varepsilon_0$ is the dielectric constant of vacuum, $k_B$ is the Boltzmann constant, T is temperature in Kelvin, n is concentration of ions, z is the valence number of these ions and e is the electronic charge. Any EDL formation inside these nominally charged surface of AAO pores are estimated to have $\lambda_{Debye}$ of the order of < 1nm only for 0.1M solutions [13] used in this synthesis. This estimate of $\lambda_{Debye}$ barely increased to ~ few nm for 0.005M of salt solutions. Here we have also assumed that these dilute salt solutions are fully ionized. We notice that these estimated values of $\lambda_{Debye}$ are much smaller compared to the average outer diameter of these CdS nanotubes (d ~ 20nm to 100nm). Therefore, we are not exactly in the traditional nanofluidics [14,15] regime where formation of EDL can effectively prevent co-ions and allow counter ions to flow inside the columnar pores. Moreover, average wall thickness of these nanotubes is also ~20nm. This is still bigger than the width of any EDL under experimental concentration used during two chamber synthesis. As a result, any typical nanofluidic contribution from EDL towards the flow of reactants will be ignored in subsequent discussions.

## B. Directionally asymmetric growth is driven by faster capillary flow of cadmium ions through AAO nanochannels.

We note that the pH of 0.1M $CdCl_2$ is at 6.5 and that of 0.1M $Na_2S$ is at 13. It is also well known [16,17] that the surface charge of AAO is determined by the pH of the solution. The iso-electronic point of AAO is generally around a pH of 8-9. Therefore, it is likely that transport [18] of chemical precursor ions through these AAO nanochannels are mostly governed by capillary flow rates which are driven by the existing chemical potential difference on two sides of an AAO nanochannel. Moreover, very high value of pH is expected to increase negative surface charge on AAO and consequently produce additional resistance to the capillary flow of $Na_2S$ based precursor at a level higher than the $CdCl_2$ based precursor solution. Whereas, with a low pH of 6.5, cadmium containing ions can face smaller resistance to its capillary flow inside these narrow nanochannels as compared to sulfur containing ions. As a result, we guess that this negatively charged walls of AAO nanochannels at such high pH levels can actually slows down the capillary flow of the $Na_2S$ precursors within these AAO pores as compared to the faster flow rate of $CdCl_2$.

We have done simple experiments to test the above statement by allowing – (1) 0.1M $CdCl_2$ or (2) 0.1M $Na_2S$ to flow for one hour through identical AAO nanotemplate having pore diameter around 100nm in two separate experiments. In both cases, (1) 100ml of 0.1M $CdCl_2$ or (2) 100ml of 0.1M $Na_2S$ is kept on one side of the AAO membrane while the other side of the cell [Fig. 1(a)] is filled with 100 ml of 18.2 MΩ de-ionized water only. Trace amounts of $CdCl_2$ ($Na_2S$) then flows through AAO nanopores and finally reach the de-ionized water bath. After one hour of flow, additional (2ml) 0.05M $Na_2S$ (0.05M $CdCl_2$) is mixed with 6ml of de-ionized water bath solution containing remnant traces of $CdCl_2$ ($Na_2S$) precursors to form CdS. In each case, CdS is formed within excess $Na_2S$ or $CdCl_2$ respectively. Here, we note that the amount of such flow is negligible enough so that the pH of the de-ionized water is not changed significantly during that process. Subsequent optical absorption experiments of these CdS particles dispersed in water clearly show much higher optical absorbance [Fig. 6(a)] in the case when only $CdCl_2$ was flowing than the case when only $Na_2S$ was flowing through identical AAO membranes. We know that optical absorbance (A) is directly related with the concentration of CdS as $A = \varepsilon L c$, where ε is the molar extinction coefficient, L is the optical path length (= inner diameter of the cuvette) and c is the molar concentration of CdS. We also see from another optical absorption experiments that ε is slightly different for solid CdS prepared in two different ways, one having excess $Na_2S$ (by mixing 0.001 M $CdCl_2$ and 0.04 M $Na_2S$ ) and the other having excess $CdCl_2$ (by mixing 0.04 M $CdCl_2$ and 0.001 M $Na_2S$). From the respective absorbance spectra [Fig. 6(b), at 445 nm], we estimate ε of CdS formed within excess amount of $Na_2S$ as $8.91 \times 10^2$ /mol.cm. The value of ε for CdS formed within excess amount of $CdCl_2$ is found to be $3.8 \times 10^2$ /mol.cm. Using these values of ε and absorbance at 445nm from Fig. 6(a), we find the concentration of CdS obtained from separate flows of − (a) only $CdCl_2$ and (b) only $Na_2S$ through the porous alumina nanochannels for similar time duration. These are respectively 0.22 mM and 0.132 mM within 100ml of de-ionized water. This clearly shows that the amount of flow for 0.1M $CdCl_2$ is larger than the amount of flow of 0.1M $Na_2S$ through similar AAO nanochannels. Therefore, we think that the likely reason for this directionally asymmetric structural out growths of CdS nanotubes is the difference in effective flow resistance (e.g viscous drags) on the capillary flows of the two chemical precursors as evident by the difference in respective flow rates. Moreover, from the nominal use of Ficks law as $J = -D\frac{\partial C}{\partial x}$, where J in $Mol/cm^2.sec$ is the flux of ions as estimated above, D is the diffusion coefficient, C is in $Mol/cm^3$, $\frac{\partial C}{\partial x}$ is the concentration gradient ~ 0.1M across 30 µm length of these nanochannels in both cases. This way, we estimated the diffusion coefficients for $Cd^{++}$ and $S^{--}$ ion flowing through alumina nanochannels as $5.8 \times 10^{-7}$ $cm^2/s$ and $3.5 \times 10^{-7}$ $cm^2/s$ respectively. This difference of pH on both side is smaller for synthesis done at a lower concentration of 0.005M, as the pH of $Na_2S$ comes down to a level of ~11. As a combined effect of smaller pH difference and smaller concentration of active reagents, the growth process of CdS nanotubes also takes much longer time (~ 24 hours) for the nanotubes shown in Fig. 3(d).

Alternatively, in case of any possible bulk like diffusive flow, $Cd^{++}$ ion (ionic radii ~ 0.109 nm) can also diffuse much faster than $S^{-2}$ ions (ionic radii ~ 0.184 nm) due to its smaller size. However, under current experimental conditions, such bulk like diffusion controlled flow through these AAO nanochannels seems unlikely. Moreover, chemical precursor solutions need to intermix and react before any size selective diffusion of $Cd^{++}$ and $S^{--}$ can even start. It is also implausible that observed tube like structures can originate from such diffusion controlled bulk like flows from both sides of AAO nanochannels. In fact, such bulk diffusion and mixing of ions followed by reaction of chemical precursors within these nanochannels can even block any further flow of reactants due to the formation of solid CdS inside these AAO nanochannels as mentioned above. Therefore, we rule out such processes as a possible cause for structural formation of these CdS nanotubes.

**C. Mechanism of CdS nanotube nucleation at the edges of AAO nanochannels and the weakest structural point of these nanostructures.**

We comprehend that the very first capillary meniscus containing cadmium ions, which reaches the other end of AAO nanochannel may have a plug like profile on its exit. This is clearly visible in Fig. 4(a) and depicted in Fig. 5(a). This initial 'cap' layer of CdS forms when the meniscus of cadmium precursor solution collides with that of sulfide ion reservoir to build the observed closed-end tip like structure of these CdS nanowires. Nevertheless, AAO nanochannels can be effectively clogged by the formation of a rigid disc or cylindrical solid structure of CdS strongly attached to its inside edge. This can eventually stop any further growth of these nanotubes if these chemical reagents are physically separated from each other in the process. However, we see continued growth of these nanotubes for few hours (Figs. 4). Moreover, from the thin walled tubular structures visible [Figs. 1(b), 3(b) etc] at the broken ends of the CdS nanowires, we presume that further outgrowth of CdS nanowires can only takes place in the form of nanotubes. It is again evident that the top ends of these nanotubes are closed from the beginning. This is clearly visible from SEM images [Figs. 4] of samples grown with different time duration of nanoreactor synthesis. However, we also see that the length of these nanotubes is increasing horizontally with increasing time duration of synthesis from thirty minutes to two hours. So we perceive that the nucleation for this outgrowth has to germinate along the periphery of these nanotubes near the edges of the AAO nanochannel as the top end (tip away from the AAO surface) of these CdS nanotubes are already sealed [Fig. 4(a)]. As a confirmation of this hypothesis, we always notice another important structural fact [Figs. 1, 3, 4] that broken edges of these nanotubes appear only on the surface of AAO template where the expected nucleation of such tubular growth actually takes place. Therefore, we perceive that the *weakest structural point* of these tubular CdS nanotubes is most likely to be at this surface of AAO nanotemplate where the above mentioned nucleation is anticipated to take place. Such active nucleation along these joints between AAO nanochannels and CdS nanotubes can certainly make it fragile and prone to mechanical failure. As a result, most of these nanotubes break only from these joints during the post-growth washing and drying sequences and also during sample processing stages for SEM imaging. Moreover, smooth outer surface of each individual nanowire also suggest that the nucleation geometry is well controlled by the rigid inner walls of these AAO nanochannels. If the same nucleation were to take place at the free ends of these nanotubes as per Fig. 5(b), then there is nothing to confine the mixing of chemical precursors and to preserve the structural uniformity of outer diameter/shape of each individual nanotubes. Therefore, this observed longitudinal growth of the 'closed-end-tip' like structure of these CdS nanotubes having uniform outer diameter clearly rules out any possible nucleation and further growth from the furthest end of these nanotubes as described in Fig. 5(b).

**D. Need for hydrodynamic instabilities for sustained nucleation of CdS nanotubes near the edges of nanochannels on the surface of AAO template.**

At this stage, it should be noted that any laminar type capillary flow of cadmium ions at the exit points of these nanochannels will not sustain the incremental growth (as seen in Fig. 4) of CdS nanotubes having such uniform outer diameter for few hours. Necessary intermixing of chemical precursors needed for the creation of CdS can easily stop if the peripheral reactive layers [Fig. 5(a)] are clogged by solid CdS and subsequently two precursor solutions can be physically separated from each other. Therefore, we emphasize the need for active presence of *interfacial convective instabilities* at the liquid-liquid ($CdCl_2$-$Na_2S$) reactive fronts around such peripheral ends of AAO nanochannels. That way, cadmium and sulfide ions can be continually driven to each other's vicinity [see Fig. 6(c)] and then react to form solid CdS whose geometry is defined by the rigid inner walls of AAO nanochannels. Having very different values of pH may also result in surface tensions gradients as well as viscosity gradients of both $CdCl_2$ and $Na_2S$ precursors around this reactive periphery in addition to any decrease of respective reagents as a result of the ongoing chemical reaction. Most importantly, chemical reaction of the fluid components may also create microscopic temperature gradients and concentration gradients of the solutes which can feed on this whole process and continually drive such instability. We will discuss this aspect in more detail in the next Sec. IV E. Under these circumstances, such gradients can change radially outwards from the center of such nanochannels and it can slowly drag $CdCl_2$ solution towards the peripheral reactive front and continue the nucleation cycle [Figs. 5(a) and 6(c)]. We argue that the presence of such convective flow near the peripheral edges of these AAO pores is required to efficiently drive the intermixing of these chemical precursors. This continuous convective outflow of $CdCl_2$ and mixing actions are needed to supply the required chemical reactants for such prolonged (few hours) nucleation process. Moreover, any differential shear caused by above mentioned chemo-hydrodynamic instability may form vortices around the reactive interface. We guess that a larger portion of this instability (vortices) may actually reside in the $CdCl_2$ side of this peripheral reactive interface [Fig. 6(c)] due to its comparatively lower viscosity and smaller pH (Sec. IV B). The wall thickness of these CdS nanotubes within the AAO pores may also be related to some characteristic size of these vortices over which chemical precursors overlap, intermix and finally react. We also expect that growth of CdS nanotubes is sustained due to this continuous nucleation process mediated by hydrodynamical instabilities at the interface of this *'tertiary fluid'* composed of – (i) $Na_2S$ and (ii) $CdCl_2$ precursor solutions and also (iii) the CdS particulates in solution phase. However, one has to investigate the quantitative details and origin of such hydrodynamical instability at such small length scales.

**E. Conventional description of hydrodynamic instabilities may not be sufficient to explain the occurrence of instabilities at the nanoscale and further challenges.**

To start with, we tend to ignore any role of buoyancy induced Rayleigh [19,20] type instabilities as CdS nanotube walls grow more or less uniformly on all sides during its horizontal growth process. Within the experimental spatial resolution of our SEM imaging, these CdS nanotubes also grow without any visible undulation of wall thickness. Subsequently, we like to explore whether Benard-Marangoni [21] type of instabilities at this peripheral reactive interface as a plausible cause for nucleation of CdS nanotubes. In order to examine these issues, we notice that relative insensitivity of the wall thickness of these nanotubes on the molar concentration of chemical precursors [Figs. 3(b) and 3(d)] may be an indication that temperature difference induced instability is playing a major role than the concentration difference induced solutal instability. We also perceive that the local resource of such temperature gradient can easily come from two sources – (i) the heat of formation of CdS [$\Delta H_{CdS}$ = -161.9 kJ/Mol; [22] (exothermic reaction)] as well as (ii) the heat of formation of nanotubular CdS structures from CdS particulates dispersed in solution phase. This localized exothermic heating is possibly the main perturbing force to sustain such convective instabilities which continuously nucleate the outgrowth of these CdS nanotubes over a prolonged duration. However, we note that net exothermic heat flow proportionately decrease with decreasing concentration of the reactants in solution. We also want to mention that there are reports [23-25] of similar interfacial instabilities of chemo-hydrodynamic origin. These instabilities are mainly governed by exothermic or endothermic chemical reactions. It was discussed [23] that non-equilibrium coupling between chemical reactions and hydro-dynamical fluctuations can induce such convective interfacial instability. In the current context, the reactive interface which is expected to produce CdS material can act as a 'chemical generator' to locally destabilize the fluid flow. It was also noted [23] earlier that exothermic reactions like CdS nanotube formation can easily produce such instabilities which may even results in effects similar to viscous fingering. Therefore, the primary reason for such convective interfacial instabilities can be attributed to the presence of significant chemo-hydrodynamic coupling at the nanoscale as a consequence of reactive nature of the fluid components. Moreover, both thermal instability and solutal instability can just add to this initial chemo-hydrodynamic trigger which starts the necessary mass flow across this reactive interface for sustained nucleation.

We estimate the dimensionless Marangoni number to get a better picture of such hydrodynamical instabilities required for prolonged nucleation of CdS Nanotubes. We use the classic definition of Marangoni number as

$$M_{Th} = \frac{-\left(\frac{\partial \gamma}{\partial T}\right) t \Delta T}{\mu \kappa} \ . \qquad (2)$$

Standard parameters of NaCl-water solution are used as reference to predict the onset of such convective instabilities in similar salt solutions of (e.g $CdCl_2$ etc) confined inside such small dimensions. We choose the average diameter of these AAO pores as the nominal fluid thickness (t) in these calculations. We use typical 'bulk' or macroscopic values [26] like thermal gradient of surface tension as $\sigma_T = \left(\frac{\partial \gamma}{\partial T}\right) = 0.161 \times 10^{-3}$ N/m.K for NaCl-water mixture, $\Delta T = 1K$, t = 100 nm, dynamic viscosity as $\mu = 0.798 \times 10^{-3}$ N.s/m$^2$, thermal diffusivity as $\kappa = 1.467 \times 10^{-7}$ m$^2$/s in this calculation. Estimated value of the thermal Marangoni number ($M_{Th}$) ~ 0.137 is much smaller than the usual critical value of $M_C = 80$ required for the onset of such classical fluid dynamical instabilities. Alternatively, we can also assume the extreme case of insulating boundaries and approximate that the mass of CdS by-product and the mass of the heated portion, however small, are nearly same. We use the net exothermic heat of reaction for $CdCl_2 + Na_2S = CdS + 2NaCl$ as $\Delta H_{Total}$=-226.7 kJ/mol=1.117 kJ/g using values like $\Delta H_{CdS} = -161.9 \, kJ/mol$, $\Delta H_{CdCl_2} = -391.5 \, kJ/mol$, $\Delta H_{Na_2S} = -364.8 \, kJ/mol$ and $\Delta H_{NaCl} = -411 \, kJ/mol$. Further using $\Delta Q = ms\Delta T$, we find that the value of $\Delta Q$ for exothermic chemical reaction to produce $1.4 \times 10^{-17}$g (using molar concentration of 0.1M) of CdS in the 100 nm$^3$ volume of porous alumina is $1.57 \times 10^{-14}$J, where m is the mass of water as $10^{-15}$g in the 100 nm$^3$ volume of porous alumina, specific heat (s) of surrounding water as 4.1813 J/g.K. From there, we roughly estimate $\Delta T$ around ~ 3.7K. Therefore, we get a ballpark figure of thermal Marangoni number ($M_{Th}$) of ~ 0.50 which is still smaller than $M_C = 80$. However, it is not uncommon in the literature [27] to find examples of similar convective instabilities at nanometer length scales even when the estimated Marangoni number is very small (~ $10^{-8}$). It is also possible that values of surface tension gradient, viscosity and thermal diffusivity can be strongly size dependent [28,29] at the nanoscale to start such instability. Moreover, there is no unique experimentally addressable expression in the literature for such complex Marangoni type dimensionless number triggered by chemo-hydrodynamic fluctuations so that one can make an educated guess on the onset of this required instability. Interestingly, this kind of fluid dynamical instabilities may actually depend on how surface tension gradient as well as the viscosity gradient is affected by the rate of interfacial chemical reactions. To draw an analogy, faster reaction rate in such cases may mimic a faster evaporation rate in case of Marangoni effects for the usual liquid-gas interface. In the same way, a faster reaction at the liquid-liquid interface can also result in a stronger convective perturbation. We also predict that an increase in the footprint of this peripheral intermixing region can transform a tubular growth to a solid rod like growth of CdS. Therefore, we predict that any CdS nanowire formed inside AAO nanopores with diameter below 20nm will be formed as solid nanorods rather than hollow nanotubes.

As mentioned above, it is possible that viscous fingering effect [23,30,31] at the miscible fluid-fluid reactive interface can be self-assembled and nucleated by ordered arrays of nanopores of AAO template. This can happen when one chemical reagent may preferentially diffuse through the common reactive interface into the opposite side and react with the other reagent. Therefore, it may also be a candidate [Fig. 5(c)] to initiate the structural growth of such CdS nanotubes where chemical kinetics and the difference [31,32] in diffusion coefficients of chemical species play a major role. Nevertheless, to sustain such structural growth over the course of time, we still need continued nucleation at the peripheral edges of these nanotubes on the surface of AAO nanochannels without any clogging. Therefore, we still argue that chemo-hydrodynamic instability at the peripheral joint of CdS nanotube-AAO nanochannel is the most likely driving force for sustained structural growth of these tubular nanostructures. Thermal and solutal driven chemo-hydrodynamic changes of surface tension at the active interface can also germinate additional convective instability as mentioned above. At this stage, it is however difficult to quantitatively separate the individual contributions of different convective interfacial instabilities in this multi-component fluid flow for reasons mentioned above. Modeling of such multi-component, possibly nonlinear [31, 32] fluid dynamics of interfacial convection under exothermic chemical reactions of the fluid components is expected to be challenging. It has been recently predicted [31, 32] that chemo-hydrodynamic triggers can not only assist an existing viscous fingering instability but it can also start one, depending on relative viscous parameters of the reactive fluids. Therefore, it is not unreasonable to demand that chemical kinetics (e.g the rate of chemical reaction etc) can affect both the surface tension gradient and the viscosity gradient in the vicinity of a reactive interface which subsequently germinate chemo-hydrodynamic instability at such small length scales. A viable model of nucleation and structural growth at the nano scale should also take into account of the energy cost towards the formation of energetically favorable tubular structures of CdS. Detailed quantitative understanding of such complex flow process inside confined spaces, nature of chemo-hydrodynamic convections as well as the growth mechanism of tubular formations are certainly needed here. Generic understanding of such chemo-hydrodynamic fluid flow and subsequent nano-fabrications at such small length scales will also be useful in bio-physical and bio-medical contexts.

# V. CONCLUSIONS.

In summary, we presented structural evidences on the hydrodynamic origin of two-chamber nanofabrication process where nanochannels of porous alumina were used as '*nano chemical reactors*' to synthesize ordered arrays of CdS nanotubes oriented in one direction. We suggested the presence of strong chemo-hydrodynamic coupling as a plausible cause to trigger and also to sustain such directionally asymmetric growth of nanoscale solid structures. We discussed how dissimilar flow resistance (e.g viscous drags) towards the capillary flow of chemical precursors can initiate such growth process, which finally yield directionally asymmetric outgrowth of CdS nanotubes only on one side of AAO template. We clearly demonstrated that nucleation sites of such nanotube like structural growths are located at the joints of these CdS nanotubes on the surface of AAO nanotemplate. We also argued in favor of chemo-hydrodynamically driven instability at the multi-component liquid-liquid reactive interface as the most likely cause of sustained horizontal growth of these CdS nanotubes. However, preliminary estimates expectedly imply that the usual description of hydrodynamic instabilities (e.g Marangoni effects etc) cannot account for the onset of such instabilities at the nanoscale. Therefore, all these results and analyses presented here build a strong case for new experimental investigations as well as theoretical developments in such directions. These are necessary to explore the physical origin of multi-component reactive fluid flow through nanochannels and also to examine the mechanism of chemo-hydrodynamic instabilities at the nanoscale in more detail.

## ACKNOWLEDGEMENTS

Authors want to thank Dept. of Science and Technology, India for DST Nano Unit grant SR/NM/NS-42/2009 and IISER-Pune for the support. We also thank National Chemical Laboratory (NCL), Pune, India for SEM images. SD wishes to thank Prof. K. N. Ganesh and IISER-Pune for startup funding. We also thank Dr. Apratim Chatterji and Dr. Arijit Bhattacharyay for valuable discussions.

## AUTHOR INFORMATION

Email(s) of corresponding author(s) : *arthurvarghese@gmail.com, **shouvik@iiserpune.ac.in

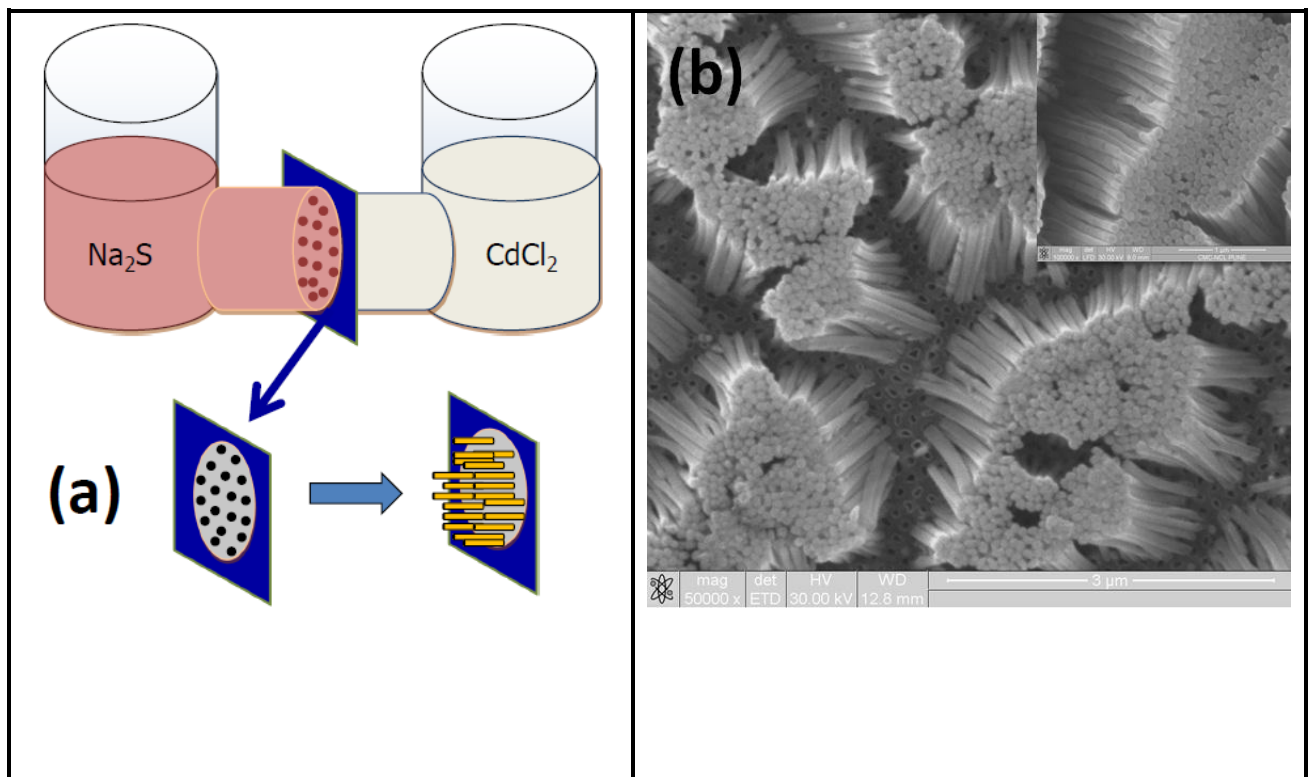

FIG. 1. (Color online) (a) Schematic diagram of two chamber *nanoreactor* synthesis of CdS nanotubes. Nano-porous AAO membranes (black dotted) are fabricated and held inside a rigid Aluminum foil (Blue). Two reagents can mix only through the porous nano-channels of AAO. The diagram also depicts the formation of nanowires (marked yellow) only on one side of the AAO template. (b) SEM surface topography of CdS nano-wires (diameter ~ 100nm) growing horizontally from one side of the AAO membrane facing the $Na_2S$ solution. Some of these nanowires are dislodged from the AAO surface during post-growth washing and drying sequence and in some other places these are bunched together. The inset shows a magnified image of one such portion. AAO template was anodized in 0.3M Oxalic acid at 60V.

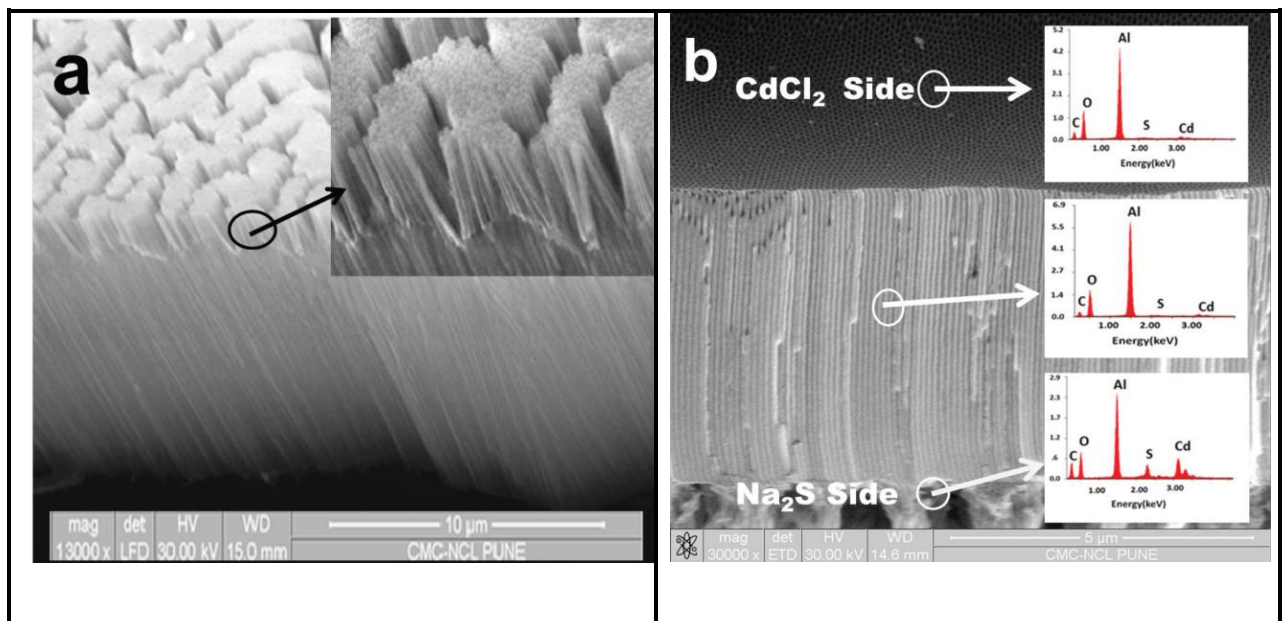

**FIG. 2.** (a) Cross sectional SEM image of as grown CdS nanowires on the top side of AAO template (anodization voltage 60V) facing Na$_2$S during the two-chamber nanoreactor synthesis process. Magnified portions in the inset is showing ~1.5μm long CdS nanowires attached to the AAO template. (b) SEM image shows the clear absence of nanowire outgrowth on the other side of AAO (anodization voltage 40V) facing CdCl$_2$ solution. EDS spectra taken from various parts of this AAO template do not show any significant presence of both 'Cd' and 'S' atoms on the CdCl$_2$ side of the template. Circles and arrows indicate the portion of the AAO template used for cross-sectional EDS scan.

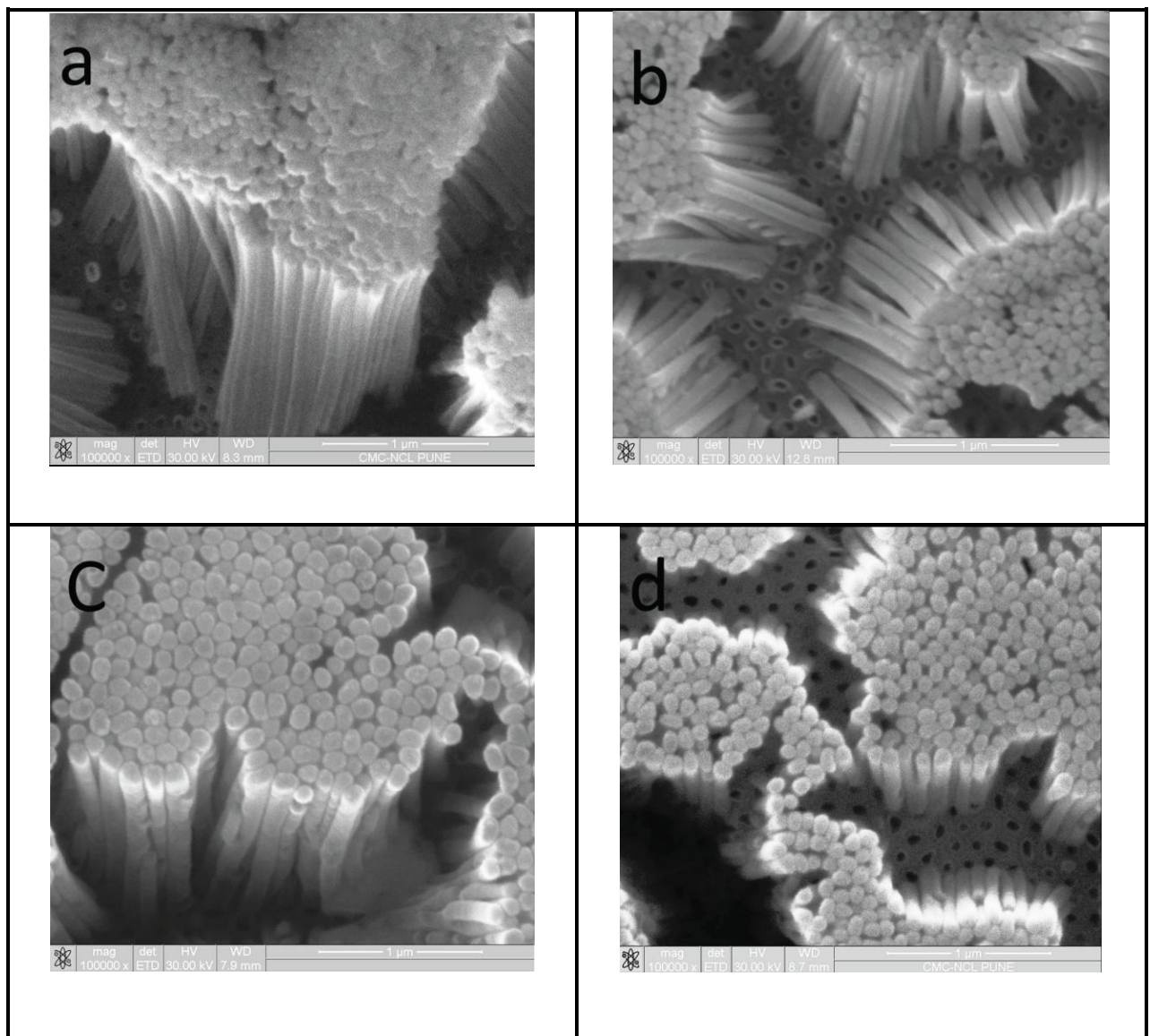

**FIG. 3.** SEM images of CdS nanotubes grown inside AAO pores fabricated in 0.3M oxalic acid under anodization voltage of (a) 50V, (b) 60V and (c) 70V and further synthesized with 0.1M $CdCl_2$ and 0.1M $Na_2S$. (d) CdS grown on 60V anodized AAO by 0.005M $CdCl_2$ and 0.005M $Na_2S$ precursors. The top end of each of these nano-tubes is also closed. We also note that the tube-like edges of broken nanotubes are lying on the surface of AAO nanotemplate. These are clearly the weakest structural points of CdS nanotubes. Therefore, we guess that these CdS nanotubes are probably nucleated around such peripheral exit points of the AAO nanochannels. These arguments are further substantiated by the results shown in Figs. 4.

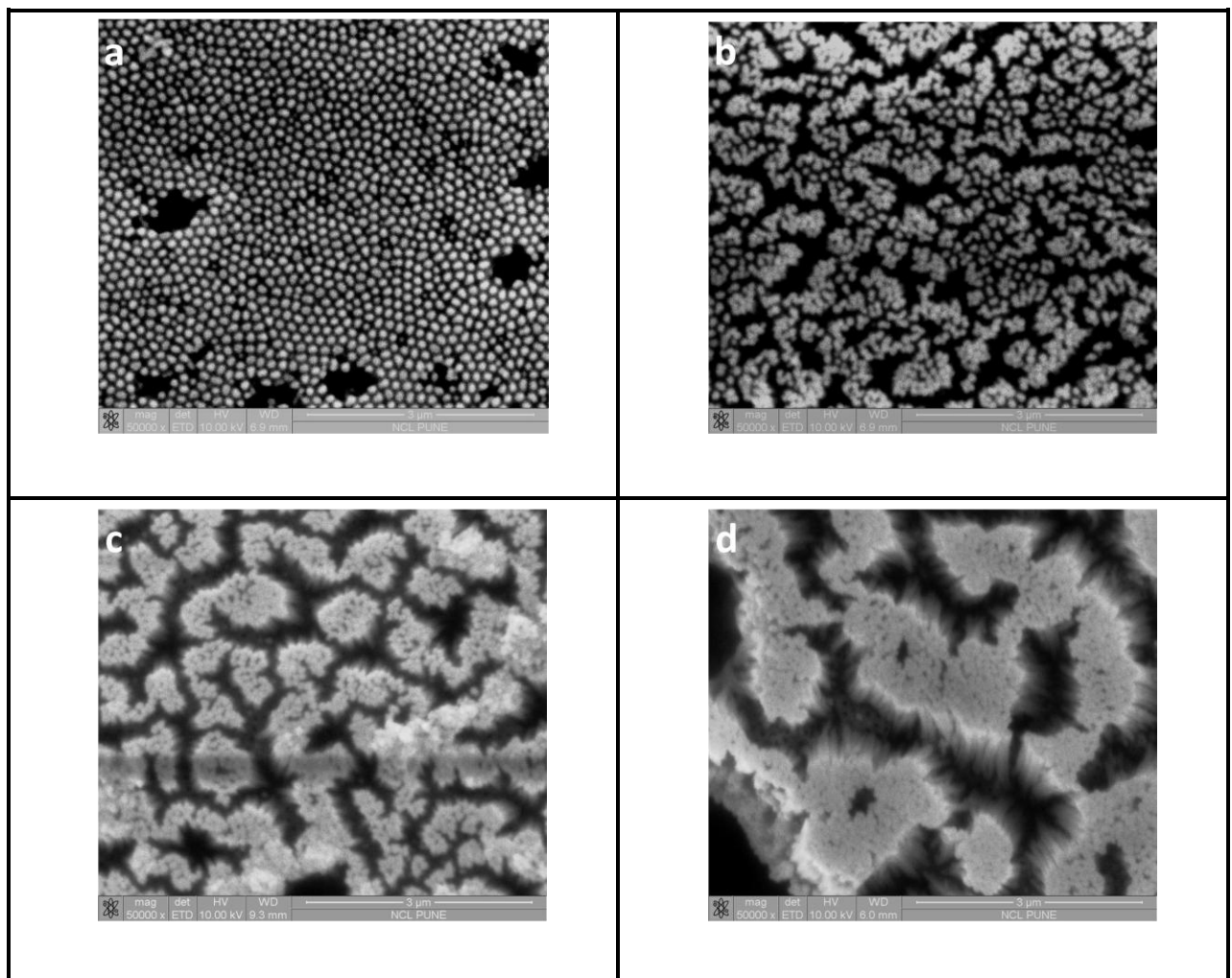

**FIG. 4.** SEM images of CdS nanotubes synthesized inside AAO pores of diameter ~100nm for different time durations - (a) 30 minutes, (b) 1 hour, (c) 1.5 hours, (d) 2 hours. Top ends of all these nanotubes are closed. However, these nanotubes are still growing in length from 30 minutes to 2 hours of synthesis durations. These results certainly supported our claims about the possibility of growth nucleation at the peripheral joint of CdS nanotubes on the surface of AAO nantemplate. Moreover, these images clearly rule out the growth mechanism described in Fig. 5(b). The AAO template was anodized in 0.3M Oxalic acid at 60V.

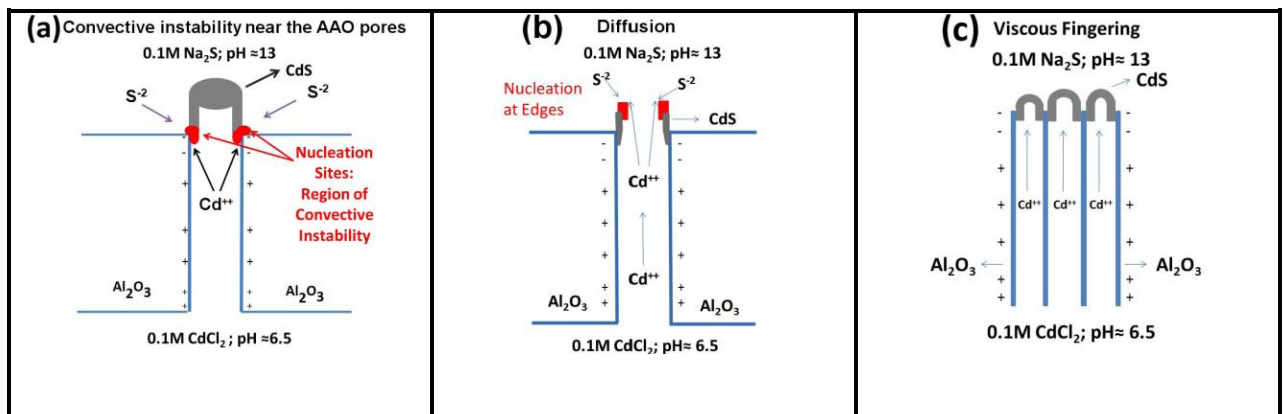

**FIG. 5.** (Color online) Comparison of three different physical mechanisms for nanotube growth. (a) Schematic diagram on the anticipated presence of convective interfacial instabilities near the peripheral end of AAO nanochannels to nucleate the tubular growth process of CdS nanowires. (b) Diffusion controlled flow of $Cd^{++}$ through the open ends of the nanotubes and possible growth nucleation at the edges of these open ends. (c) Growth of CdS nanotubes due to self assembled viscous fingering effect nucleated by the ordered nanopores of AAO. We rule out the second mechanism of selective diffusion of $Cd^{++}$ as we do not see open ended CdS nanotubes at any stages of the growth progression [Figs. 4]. Actual growth process may be a combination of both Figs. 5(a) and 5(c).

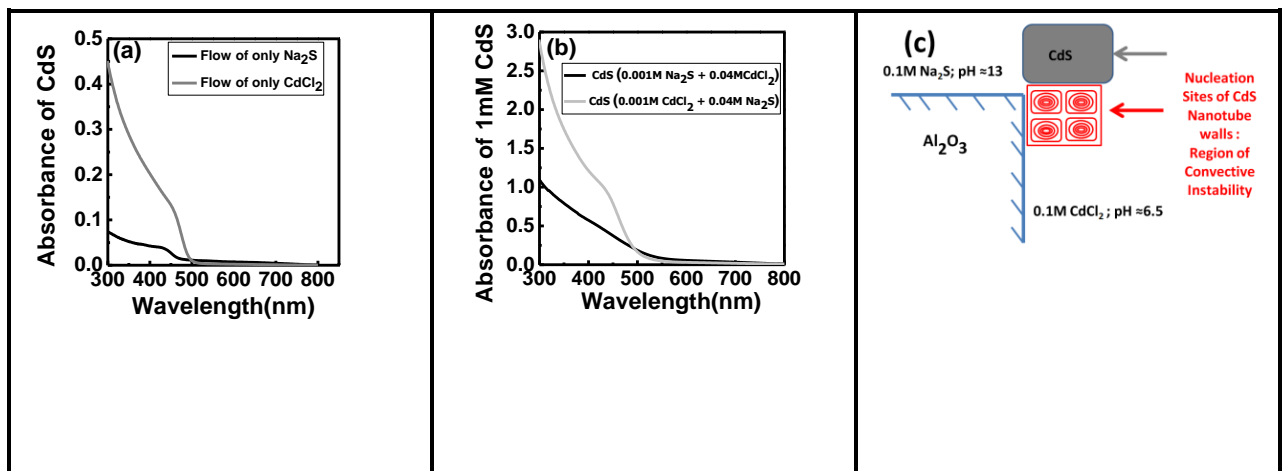

**FIG. 6.** (Color online) (a) Optical absorption spectra showing higher concentration of CdS formed due to faster flow of $CdCl_2$ as compared to $Na_2S$ through AAO nanopores resulting in larger amount of available reactants. (b) Optical absorption spectra to determine the molar extinction ratio of CdS formed by bulk mixing of $CdCl_2$ and $Na_2S$ in respective cases. (c) Representative diagram of suggested convective interfacial instability at the liquid-liquid reactive interface consisting of $Na_2S$, $CdCl_2$ solutions and CdS particulates. A large portion of this instability may reside inside $CdCl_2$ precursor solution because of its low viscosity. Sustained nucleation of CdS nanotubes requires that the intermixing of the reactants is sustained by such convective interfacial instability.

# References

[1] H. Masuda, F. Hasegwa and S. Ono, J. Electrochem. Soc. **144**, L127 (1997).

[2] D. G. Sulka, in *Nanostructured Materials in Electrochemistry*, edited by A. Eftekhari (Wiley –VCH Verlag Gmbh & Co, Weinheim, 2008 ).

[3] T. S. Kang, A. P.Smith, B. E. Taylor and M. F. Durstock, Nano Lett. **9** , 601-606 (2009).

[4] Y. Guo, Q. Tang, H. Liu, Y. Zhang, Y. Li, W. Hu, H. Wang and D. Zhu, J. Am. Chem. Soc. **130** , 9198-9199 (2008).

[5] Z. Fan, H. Razavi, J. W. Do, A. Moriwaki, O. Ergen, Y. L. Chueh, P. W. Leu, J. C. Ho, T. Takahashi, L. A. Reichertz, S. Neale, K. Yu, M. Wu, J. W. Ager and A. Javey, Nat. Mater. **8**, 648 (2009).

[6] See Online Supplemental Material (http://link.aps.org/supplemental/10.1103/PhysRevE.85.056104) for further details of synthesis and fabrication procedures of AAO nanotemplates and CdS nanotubes and for information about characterization techniques used in the work etc.

[7] F. Zang and S. S. Wong, Chem. Mater. **21** , 4541 (2009).

[8] Y. Piao, H. Lim, J. Y. Chang, W. Y. Lee and H. Kim, Electrochimica Acta. **50**, 2997 (2005).

[9] S. Kumar, J. Mater Sci. Mater. Electron. **22**, 244 (2011).

[10] Y. Lv, W. Liu, J. Huang and L. Yao, J. Nanomaterials. 546715 (2011).

[11] T. Zhai, X. Fang, Y. Bando, Q. Liao, X. Xu, H. Zeng, Y. Ma, J. Yao and D. Golberg, ACS Nano. **3**, 949 (2009).

[12] P. Abgrall and N. T. Nguyen, *Nanofluidics* (Artech House, Boston, 2009), p. 9-69.

[13] R. B. Schoch, J. Han and P. Renaud, Rev. Mod. Phys. **80**, 839 (2008).

[14] H. Daiguji, Chem. Soc. Rev. **39**, 901 (2010).

[15] J. T. Eijkel and A. Van den berg, Chem. Soc. Rev. **39**, 957 (2010).

[16] R. Yan, W. Liang, R. Fan and P. Yang, Nano Lett. **9** , 3820 (2009).

[17] E. A. Bluhm, E. Bauer, R. M. Chamberlin, K. D. Abney, J. S. Young and G. D. Jarvinen, Langmuir. **15**, 8668 (1999).

[18] T. M. Squires and S. R. Quake, Rev. Mod. Phys. **77**, 977 (2005).

[19] J. T. Chen, M. Zhang and T. P. Russell. Nano lett. **7**, 183 (2007).

[20] Y. Qin, S. M. Lee, A. Pan, U. Gosele and M. Knez, Nano Lett. **8**, 114 (2008).

[21] T. E. Faber, *Fluid Dynamics for Physicists* (Cambridge University Press, Cambridge, 2004) p. 289-342.

[22] D. R. Lide, *CRC Handbook of Chemistry and Physics* (CRC Press, Boca Raton, 2005).

[23] S. Swernath and S. Pushpavanam, J. Chem. Phys. **127**, 204701 (2007).

[24] D. A. Bratsun and A. De Wit, Phys. Fluids. **16**, 1082 (2004).

[25] S. Kalliadasis, J. Yang and A. De Wit, Phys. Fluids. **16**, 1395 (2004).

[26] J. Zhang, A. Oron and R. P. Behringer, Phys. Fluids. **23**, 072102 (2011).

[27] M. Dietzel and S. M. Troian, J. Appl. Phys. **108**, 074308 (2010).

[28] S. P. Jang, J. H. Lee , K. S. Hwang and S. S. Choi, Appl. Phys. Lett. **91**, 243112 (2007).

[29] H. M. Lu and Q. Jiang, Langmuir. **21**, 779 (2005).

[30] T. Podgorski, M. C. Sostarecz, S. Zorman and A. Belmonte, Phys. Rev. E. **76**, 016202 (2007).

[31] T. Gérard and A. De Wit, Phys. Rev E. **79**, 016308 (2009).

[32] Y. Nagatsu and A. De Wit. Phys. Fluids **23**, 043103 (2011).